\documentclass[prd,aps,nofootinbib,singlecolumn]{revtex4-2}
\usepackage{amsmath,amssymb,amsfonts,color,graphicx,graphics,latexsym,placeins,epsfig,multirow}
\usepackage[toc]{appendix}
\usepackage{array}
\newcolumntype{P}[1]{>{\centering\arraybackslash}p{#1}}
\usepackage{footmisc}
\usepackage{mathrsfs}
\usepackage[compatibility=false]{caption}
\usepackage{subcaption}
\usepackage{comment}
\usepackage{bbold}
\usepackage{enumitem}
\setdescription{leftmargin=12.5pt}
\usepackage{bbold}
\usepackage{hyperref}
\usepackage{varioref}
\usepackage{color}
\usepackage{dirtytalk}
\def\a{\alpha}

\def\6{\partial}

\begin{document}

\title{\Large \bf Sengupta Transformations and  Carrollian Relativistic Theory}
\author{Rabin Banerjee$^1$ \footnote{DAE Raja Ramanna fellow}}
\author{Soumya Bhattacharya$^1$}
\author{Bibhas Ranjan Majhi$^2$}
\affiliation{$^1$Department of Astrophysics and High Energy Physics, S.N. Bose National Center for Basic Sciences, Kolkata 700106, India}
\email{rabin@bose.res.in}
\email{soumya557@bose.res.in}
\affiliation{$^2$ Department of Physics, Indian Institute of Technology Guwahati, Guwahati
781039, Assam, India}
\email{bibhas.majhi@iitg.ac.in}

\vskip .2in
\begin{abstract}
     %In this paper, we formulate, in a systematic manner, Carrollian relativistic action in detail. Exploiting the transformations, first provided by Sengupta \cite{Sengupta}, we construct a mapping  between Lorentz relativistic and Carrollian relativistic vectors. 
     A detailed and systematic formulation of Carrollian relativity is provided. Based on the transformations, first provided by Sengupta \cite{Sengupta}, we construct a mapping  between Lorentz relativistic and Carrollian relativistic vectors. Using this map the Carroll theory is built from the standard Maxwell action. We show that we get self-consistent equations of motion from the action, both in electric and magnetic limits. We  introduce Carroll electric and magnetic fields. A new set of maps is  derived that connects Carroll electric and magnetic fields with the usual Maxwell ones and yields Carroll equations in terms of fields. Consistency of results with the potential formulation is shown. Carroll version of symmetries like duality, gauge, shift, Noether and boost are treated in details and their implications elaborated. Especially, boost symmetry provides a link to the various maps used in this paper.
\end{abstract}  
\maketitle
\section{Introduction}
We know that the non-relativistic limit of the relativistic physics plays an important role in modern physics. These non-relativistic theories have found applications in diverse fields \cite{Taylor, andreev1, andreev2, jensen, rb1, rb2, Pal, Jain, rb3, Morand}. The non-relativistic limit can be obtained by group contraction method from Poincare group which describes the relativistic sector. One of the well studied examples of non-relativistic limit is the Galilean limit where one considers the speed of light to be considerably greater than the other speeds. The study of Galilean relativistic theories was initiated by Le Bellac and Levy Leblond \cite{Leblond} back in 1970’s. Afterwards various aspects of Galilean relativistic physics have been studied \cite{Khanna, Duval, Bhattacharya, Bhattacharya1, Bhattacharya2}. A very good discussion on the physical implications of Galilean electrodynamics can be found in the review article \cite{Rousseaux1}.\footnote {Galilean aspect of Coulomb and Lorentz gauge has been discussed in \cite{Rousseaux3} and the usage of Galilean equations in the context of momentum transfer is shown in \cite{Rousseaux4}}.
 
Another unfamiliar yet quite interesting aspect to study occurs when the speed of light is considerably smaller than other speeds. This limiting case was first proposed by Levy-Leblond \cite{CLeblond} and almost simultaneously by N. D. Sengupta \cite{Sengupta} although the methods were quite different. Levy-Leblond \cite{CLeblond} coined the name `Carroll group' for the corresponding symmetry group, by referring to Lewis Carroll's books, {\it ``Alice in the Wonderland"} and {\it ``Through the Looking Glass"}, for showing a strange world (breakdown of causality, for instance) (for further details look at \cite{Duval}). Contrary to the Galilean case, where light cones
open up, the light cones close up in the Carroll limit. This has some unusual consequences
for the kinematics and dynamics in this limit. Immediately after discovery of the Carroll group  interest was only limited to mathematical curiosity and the field otherwise lay fallow due to lack of solid physical motivations. But in the last few years we see a resurgence of interest in this field. One of the main reasons behind this is the identification of the conformal extension of the Carroll group and the Bondi-Metzner-Sachs (BMS) group \cite{Bondi} which plays important role in gravitational physics \cite{Duval2}. There have been a lot of studies on Carrollian sector in various fields like basic formalism \cite{Duval}, in gravitational waves and memory effects \cite{Duval3, Duval4}, in the Hall effect \cite{Horvathy}, in quadratic and higher derivative gravity theories \cite{Tadros1, Tadros2}, in the hydrodynamics \cite{Petkou, Petkou2}, in the celestial holography \cite{Donnay, Bagchi} to name a few. 

Here in this paper we are mostly interested in the formulation aspect of the Carroll relativistic field theory. Almost all the work in this field follow the formalism put forward by Levy-Leblond \cite{CLeblond} which is based on the argument of the principle of causality. For example a Hamiltonian formulation of Carrollian relativistic theory is provided in \cite{Henneaux}. We, on the other hand are more interested by the kinematic approach proposed by N. D. Sengupta \cite{Sengupta}. He gave a new set of transformations from which we are able to derive known Carroll relativistic results. However Sengupta's work \cite{Sengupta} received little recognition \footnote{Very recently Levy-Leblond quite decently mentioned about Sengupta's work in a review \cite{Leblond2}}. Our main motivation here is to build up a systematic action formalism for a Carroll relativistic field theory following the similar track that we have used for the Galilean relativistic field theories \cite{Bhattacharya, Bhattacharya1, Bhattacharya2} using Sengupta's transformations. One cannot get the desired results by starting with usual Lorentz transformations simply because they are not valid. Here Sengupta's formulation plays a crucial role. We show in our paper in a very straight forward way how to build up a consistent Carroll relativistic action formulation by using Sengupta's formulation \cite{Sengupta}. 

We begin our paper by introducing briefly the Lorentz transformations and the Galilean limits. Here we discuss about the mappings that connect the Lorentz and Galilean relativistic quantities. We then move to discuss the Sengupta's transformations. Keeping in mind the Galilean analogue we  provide a mapping  connecting relativistic and Carroll vectors for both `electric' and `magnetic' limits. Using these  relations we then build up the action formulations in terms of potentials for both these limits and write down explicitly the equations of motion and their internal consistency. We then introduce the Carrollian electric and magnetic fields and the entire analysis is then performed in terms of electric and magnetic fields. Along the way, we discuss the duality symmetry involved here. Next we  discuss gauge symmetry. We also compute the Carrollian version of the Noether currents and explicitly show their on-shell conservation. Shift symmetries which play an important role in the study of low-energy effective Lagrangians in the context of Goldstone’s theorem  are analysed from a modern viewpoint. Recently there have been some studies on such symmetries from different aspects \cite{Bhattacharya, rabin1, rabin2}. We compute corresponding currents and their conservations in the Carroll limit.

The paper is organised as follows. In section \ref{sec2} we provide a brief review of the Galilean limit and the mapping relations. Section \ref{sec3} provides a short   discussion about Sengupta's formulation and the Carrollian mappings. In section \ref{sec4} we build up a systematic action formulation for Carroll relativistic Maxwell theory for both electric and magnetic limits. In section \ref{sec5} we introduce the Carrollian electric and magnetic fields and recast the equation of motion in terms of fields. Transformations under Carrollian boost have been discussed in section \ref{sec6}. Gauge symmetry, Noether currents and their conservations are discussed in section \ref{sec7}.  In section \ref{sec8} we discuss shift symmetry and its Carrollian counterpart, corresponding currents and their conservations. Finally, conclusions have been given in section \ref{sec9}.

\section{Lorentz transformation and Galilean limit} \label{sec2}
Here we derive a certain scaling between special relativistic and Galilean relativistic quantities. As we know there exist two types of such limits, namely electric and magnetic limits. Let us first consider the contravariant vectors. Take a generic Lorentz transformation with the boost velocity as $u^i$:
\begin{equation}
    x'^0 = \gamma x^0 - \frac{\gamma u_i}{c} x^i~;
    \label{lt1}
\end{equation}
\begin{equation}
    x'^i = x^i - \frac{\gamma u^i}{c}x^0 + (\gamma -1 )\frac{u^i u_j}{u^2}x^j~,
    \label{lt2}
\end{equation}
where the boost factor is given by 
\begin{equation}
    \gamma = \frac{1}{\sqrt{1-\frac{u^2}{c^2}}}~. 
    \label{LOF}
\end{equation}
%\textcolor{red}{Under such Lorentz transformations a contravariant vector changes as
%\begin{equation*}
%    V'^{\mu} = \frac{\6 x'^{\mu}}{\6 x^{\nu}} V^{\nu}
%\end{equation*}}
Under such Lorentz transformations contravariant components of a vector transform as (also considering $u<<c$, so $\gamma \to 1$)
\begin{equation}
    V'^0 = V^0 - \frac{u_j}{c}V^j~;
    \label{gcontra1}
\end{equation}
\begin{equation}
    V'^i = V^i - \frac{u^i}{c} V^0~.
    \label{gcontra2}
\end{equation}
%\noindent \textcolor{red}{ At this point taking the non-relativistic limit is rather tricky. A naive $c \to \infty$ limit in (\ref{lt1}) and (\ref{lt2}) does not make sense. Also, this limit is counterfactual. Thus something else has to be done. The genesis of the solution lies in the two separate components, $V^0$ and $V^i$ and comparing their relative magnitudes.} 
%\textcolor{red}{next}

We now provide a map that relates the Lorentz vectors with their Galilean counterparts and provide the `large time-like' vectors and is called `electric limit' \footnote{For detailed construction one can look at \cite{Bhattacharya}}
\footnote{Notation: Here relativistic vectors are denoted by capital letters ($V^0, V^i$ etc) and Galilean vectors are denoted by lowercase letters ($v^0, v^i$ etc).} 
\begin{equation}
    V^0 = c v^0, \,\,\,\, V^i = v^i~.
    \label{gcontrael}
\end{equation}
%\textcolor{red}{This particular map corresponds to the case $\frac{V^0}{V^i} = c ~\frac{v^0}{v^i}$ in the $c \to \infty$ limit. This yields largely timelike vectors and is called `electric limit'.}
Now using (\ref{gcontrael}) in (\ref{gcontra1}) and (\ref{gcontra2}) we get
\begin{equation}
v'^0 = v^0~;
\label{gv1}
\end{equation}
\begin{equation}
    v'^i = v^i - u^i v^0~.
    \label{gv2}
\end{equation}
We next consider a different scaling,
\begin{equation}
    V^0 = -\frac{v^0}{c}, \,\,\,\, V^i = v^i~.
    \label{gcontramag}
\end{equation}
This map corresponds to `large spacelike vectors' and the corresponding transformations are given by,
\begin{equation}
    v'^0 = v^0 + u_j v^j~;
    \label{gv3}
\end{equation}
\begin{equation}
v'^i = v^i~.
    \label{gv4}
\end{equation}
Both pairs (\ref{gv1}, \ref{gv2}) and (\ref{gv3}, \ref{gv4}) are non-relativistic transformations (since the Lorentz factor \ref{LOF} has disappeared). %\textcolor{red}{However both satisfy the group composition law. To see this take the set (\ref{gv1}, \ref{gv2}). Thus consider a transformation from the primed to a double-primed frame,
%\begin{eqnarray}
%    v''^0 = v'^0 = v^0, \label{A}\\
%    v''^i = v'^i -u'^i v'^0 = v^i - u^i v^0 - u'^i v^0 \nonumber \\
%    = v^i - (u^i + u'^i) v^0 \label{B}
%\end{eqnarray}
%where (\ref{gv1}, \ref{gv2}) has been used to obtain the final composition law. From (\ref{A}, \ref{B}) its is seen that the group composition law holds with the new velocity given by $u^i + u'^i$. Likewise (\ref{gv3}, \ref{gv4}) yields, 
%\begin{eqnarray}
 %    v''^i = v'^i = v^i, \label{C}\\
 %   v''^0 = v^0 - (u_j + u'_j) v^j \label{D}
%\end{eqnarray}}
It is useful to note that the roles of the time and space coordinates are interchanged and is a manifestation of the two limits of the theory.

It is clearly seen from the Lorentz transformations given in (\ref{lt1}) and (\ref{lt2}) that considering Carroll limit i.e $c \to 0$ limit is quite tricky. In the following section we discuss a different set of transformations which are complementary to the usual Lorentz transformations and for which considering Carroll limit is much more easier and direct.

%\textcolor{red}{Can we make this section such that it will motivate to get the Carroll transformations? e.g. give only the main idea of constructing the Galilean results so that the same ideas can be followed for the present transformation. So the derivation procedures can be summarised without any explicit calculation and the final results can be mentioned. Also a few comments on the covariant components can be made.}

\section{Sengupta's transformation and Carroll limit} \label{sec3}
\noindent Lorentz transformations are a natural route to study the Galilean limit of a theory. There are two limits - electric and magnetic. Since the electric limit corresponds to large timelike vectors it occurs meaningfully in coordinate vectors in Lorentz relativistic theories. While this is essential for presenting causality, an arbitrary four vector (say, a four potential) is not restricted by this principle of causality \cite{CLeblond}. It may be either space-like or time-like. Thus, although the electric limit is favoured, both limits may be developed using Lorentz transformations. Now arises a question. If we can develop both limits of a non-relativistic theory from Lorentz transformations, is this possible for transformations where the magnetic limit is favoured? The answer is yes, as elaborated below.   

Let us take a new set of transformations first proposed by N.D. Sengupta with the boost velocity as $\omega$ along $x$-axis. For this specific choice of boost direction these are given by \cite{Sengupta}:
\begin{eqnarray}
&&x' = \tilde \gamma \Big(x - \tilde \beta x^0 \Big)~; \,\,\,\ y'=y~; \,\,\ z'=z~;
    \label{tn1}
\\
    &&x'^0 = \tilde \gamma \Big(x^0 - \tilde \beta x  \Big)~,
    \label{tn2}
\end{eqnarray}
where the Sengupta factors are defined as, 
\begin{equation}
    \tilde \gamma = \frac{1}{\sqrt{1 - \tilde \beta^2}}~; \,\,\,\ \tilde \beta = \frac{c}{\omega}~.
    \label{BRM2}
\end{equation}
However for the present analysis we need to generalise the above ones for the boost along an arbitrary direction. This can be done very easily. Note that the position vector $\vec{x}$ can be decomposed into two parts -- one is parallel to the boost (call as $\vec{x}_{||}$) and another is perpendicular to this (call as $\vec{x}_\perp$);  i.e. $\vec{x} = \vec{x}_{||} + \vec{x}_\perp$. The parallel component must be given by $\vec{x}_{||} = \frac{\vec{\tilde{\beta}} (\vec{\tilde{\beta}}\cdot \vec{x})}{\tilde{\beta}^2}$, where we have $\tilde \beta^i = \frac{c \omega^i}{\omega^2}$ and $\tilde{\beta}^2 = \tilde{\beta}_i\tilde{\beta}^i$. Following the structure of the equation in (\ref{tn1}), the transformation law for the parallel and perpendicular components can be given as
\begin{equation}
\vec{x'}_{||} = \tilde{\gamma}\Big(\vec{x}_{||} - \vec{\tilde{\beta}} x^0\Big)~; \,\,\,\ \vec{x'}_\perp = \vec{x}_\perp~.
\label{BRM1}
\end{equation}
Here $\tilde{\gamma}$ takes the similar form as given by the first equation in (\ref{BRM2}), but now $\tilde{\beta}^2$ is defined as $\tilde{\beta}^2 = \tilde{\beta}_i\tilde{\beta}^i$.
Then the transformation law for the position vector can be obtained as
\begin{eqnarray}
 \vec{x'} &=& \vec{x'}_\perp + \vec{x'}_{||}
 \nonumber
 \\
&=& \vec{x}_\perp + \tilde{\gamma}\Big(\vec{x}_{||} - \vec{\tilde{\beta}} x^0\Big)
\nonumber
\\
&=& \vec{x} - \vec{x}_{||} + \tilde{\gamma}\Big(\vec{x}_{||} - \vec{\tilde{\beta}} x^0\Big)~,
\end{eqnarray}
where in the second equality (\ref{BRM1}) has been used. Next using $\vec{x}_{||} = \frac{\vec{\tilde{\beta}} (\vec{\tilde{\beta}}\cdot \vec{x})}{\tilde{\beta}^2}$ we obtain the desired result
\begin{equation}
    x'^i = x^i + \frac{\tilde{\gamma} - 1}{\tilde{\beta}^2} \Big(\tilde{\beta}^i \tilde{\beta}_j\Big)x^j  - \tilde{\gamma} \tilde{\beta}^i x^0~.
    \label{t1}
\end{equation}    
While (\ref{tn2}) suggests that the temporal coordinate should have the following transformation: 
\begin{equation}
 x'^0 = \tilde \gamma \Big(x^0 - \tilde \beta_j x^j  \Big)~.
    \label{t2}  
\end{equation}
%\begin{eqnarray}
%    &&x'^i = \tilde \gamma \Big(x^i - \tilde \beta^i x^0 \Big)~;
%    \label{t1}
%\\
%    &&x'^0 = \tilde \gamma \Big(x^0 - \tilde \beta_j x^j  \Big)~,
%    \label{t2}
%\end{eqnarray}
Although both $u$ and $\omega$ have dimension of velocity they have very different physical interpretations.
On the other hand, from (\ref{t2}) we find that if $\Delta t' = 0$ then $ \omega^i = \frac{dx^i}{dt}$. This implies that $\omega^i$ is the rate of motion of an event measured in the unprimed frame that occurs at a fixed instant of time in the primed frame. It should also be noted that the usual Lorentz transformations and Sengupta's transformations are related by $u^i \to \frac{c^2\omega^i}{\omega^2}$ or equivalently, $\omega^i \to \frac{c^2 u^i}{u^2}$, where $u^i$ is the Lorentz boost velocity. Recall that both $u^i$ and $\omega^i$ have dimensions of velocity. For covariant components this is generalised as, 
\begin{equation}
    u_i \to \frac{c^2 \omega_i}{\omega^2}~;\,\,\,\ \omega_i \to \frac{c^2 u_i}{u^2}~.
\end{equation}
The other point is that $u_i$ and $\omega_i$ may be regarded as independent with complementary range of values, $0 \leq |u| < c$ and $|\omega| > c$. In this way, both small and large velocity limits may be implemented. 

Using the transformations (\ref{t1}) and (\ref{t2}) we will now build the backbone of our analysis which consists of the transformation rules of an arbitrary four vector in the Carroll limit.  
Under these transformations, it is very easy to show that the contravariant components of any four-vector transform as 
\begin{eqnarray}
    V'^0 = \tilde \gamma \Big(V^0 - \tilde \beta_j V^j \Big)~;
    \label{B1}
    \\
    V'^i = V^i + \Big(\tilde \gamma - 1\Big) \frac{\tilde \beta^i \tilde \beta_j}{\tilde \beta^2} V^j - \tilde \gamma \tilde \beta^i V^0~.
    \label{B2}
\end{eqnarray}
It may be noted that within the Sengupta's formalism the Carroll limit is achieved through the following conditions: 
\begin{equation}
    c << \omega~; \,\,\,\ \tilde \gamma \to 1~; \,\,\,\ \tilde \beta^i = \frac{c \omega^i}{\omega^2} << 1~.
    \label{B5}
\end{equation}
Imposition of the above on (\ref{B1}) and (\ref{B2}) then yields 
\begin{eqnarray}
    &&V'^0 = V^0 - \frac{c \omega_j}{\omega^2} V^j~,
    \label{B3}
    \\
    &&V'^i = V^i - \frac{c \omega^i}{\omega^2} V^0~. \label{tr1}
\end{eqnarray}
The above ones will provide us the transformation rules of Carrollian vectors from one frame to other. In order to find those
we next provide a map that relates the Lorentz vectors with their Carrollian counterparts. 
Carrollian vectors $v^0, ~v^i$ are introduced by the scaling,\footnote{Notation: Here relativistic vectors are denoted by capital letters ($V^0, ~V^i$ etc.) and Carrollian vectors are denoted by lowercase letters
($v^0, ~v^i$ etc)}
\begin{equation}
    V^0 = v^0~; \,\,\,\, V^i = c v^i~.
    \label{contrael}
\end{equation}
This particular map corresponds to the case $\frac{V^0}{V^i} = \frac{1}{c} ~\frac{v^0}{v^i}$ such that $c \to 0$ limit this yields largely timelike vectors and is called the `electric limit'. In this particular limit the using the scaling relation (\ref{contrael}) the transformation equations (\ref{B3}) and \ref{tr1} will take the following forms in the $c \to 0$ limit,
\begin{eqnarray}
    v'^0 = v^0~; \nonumber \\
    v'^i = v^i - \frac{\omega^i}{\omega^2} v^0~.
\end{eqnarray}
In a similar fashion one can define the large spacelike vectors by the following scaling relation
\begin{equation}
   V^0 = c v^0~; \,\,\,\, V^i = v^i~, 
   \label{contramag}
\end{equation}
which in the $c \to 0$ limit  yields largely spacelike vectors and is called the `magnetic limit'. Now using these scaling relations (\ref{contramag}) and taking the $c \to 0$ limit we have
\begin{eqnarray}
    v'^0 = v^0 - \frac{\omega_j}{\omega^2} v^j~, \nonumber \\
    v'^i = v^i~.
\end{eqnarray}
\begin{table}
\caption{Mapping relations}\label{T1}
\begin{center}
\begin{tabular}{|c|c|c|} \hline 
${\rm Limit}$  & $ {\rm Galilean~~mapping}$ & $ {\rm Carrollian~~mapping}$  \\ \hline
${\rm Electric ~~limit}$ & $V^0 \to c~v^0, \,\,  V^i \to v^i$ & $V^0 \to v^0, \,\, V^i \to c v^i$ \\ \hline
${\rm Magnetic ~~limit}$ & $V^0 \to -\frac{v^0}{c},\,\,V^i \to v^i $ & $V^0 \to c~v^0. \,\, V^i \to v^i$ \\
\hline
\end{tabular}
\label{T1}
\end{center}
\end{table}

%\noindent \textcolor{red}{We know that in the Carroll limit the Contra and Covariant vectors represent distinct entities as they are not connected through any non-degenerate metric. One can derive the scaling relations for the covariant vectors as well by following the same track as the contravariant ones. Here in this paper we only consider the contravariant sector. }
%\noindent To represent the covariant components of four vectors we will write first the reverse transformations of eqn \ref{t1} and \ref{t2} as
%\begin{equation}
%    x^i = \tilde \gamma \Big(x'^i + \tilde \beta^i x'^0 \Big)
%    \label{t3}
%\end{equation}
%\begin{equation}
%    x^0 = \tilde \gamma \Big(x'^0 + \tilde \beta_j x'^j  \Big)
%    \label{t4}
%\end{equation}
%\noindent Similarly one can show that the covariant vectors transform under transformations \ref{t3} and \ref{t4} in the low velocity limit as 
%\begin{eqnarray}
%   V'_0 = V_0 + \frac{c \omega^j}{\omega^2} V_j, \\
%    V'_i = V_i + \frac{c \omega_i}{\omega^2} V_0
%\end{eqnarray}
%Here we choose a particular choice of scaling relations for covariant Carroll vectors as 
%\begin{equation}
%   V^0 = c v^0, \,\,\,\, V^i = v^i  
%\end{equation}
%This can be argued from the norm preservation of both relativistic and Carrollian relativistic vectors
%\begin{equation}
%    V^0 V_0 + V^i V_i \xrightarrow[\text{limit}]{\text{electric}} \Big(c v_0 \Big)\Big(\frac{v^0}{c} \Big) + \Big(v^i\Big) \Big( v_i\Big) = v^0 v_0 + v^i v_i
%\end{equation}
\section{Carroll relativistic Maxwell theory: action principle} \label{sec4}
\noindent Here we wish to systematically build up a Carrollian relativistic version of the Maxwell theory. Let us now start from the relativistic Maxwell theory described by the action given as
\begin{equation}
    S = -\frac{1}{4}\int d^4 x  F_{\mu \nu} F^{\mu \nu}~,
    \label{mlag}
\end{equation}
where $F_{\mu\nu} = \partial_\mu A_\nu - \partial_\nu A_\mu$ with $A_\mu$ is identified as the relativistic four-vector potential.
The calculation will be done following a particular general idea. Before taking $c \to 0$ limit, we will write down all the terms (if that term has any possible $c$ dependence) in any mathematical expression in terms of $c$. So relativistic four vectors can be written in terms of Carrollian vectors with possible $c$ dependence described in previous section and the partial derivatives i.e. $\partial^0, ~\partial^i, ~\partial_0, ~\partial_i$ will follow the relation: $\partial^0 = - \partial_0 = - \frac{1}{c}\partial_t$ and $\partial^i = \partial_i $. These are dictated by the relations $x^0=-x_0= c t, ~x^i= x_i $. Finally the $c\to 0$ limit will be considered. %One more important point is that we work only with the contravariant components. 
Hence we can write
\begin{equation}
    F_{\mu \nu} F^{\mu \nu} = \eta_{\mu \lambda} \eta_{\nu \rho} F^{\lambda \rho} F^{\mu \nu}= -2 (F^{0i})^2 + (F^{ij})^2~.
    \label{B6}
\end{equation}
Here $\eta_{\mu \nu}$ is the flat space metric with signature $\Big(-,+,+,+\Big)$ and we use the convention that 
%\begin{equation}
$   (F^{0i})^2 = F^{0i} \times F^{0i}$ and $(F^{ij})^2 = F^{ij} \times F^{ij}$.
%   \end{equation}
\subsection{Electric limit}
\noindent In this limit the Maxwell action \ref{mlag} will be reduced to 
\begin{equation}
    S_e = \int d^4 x \mathcal{L}_e~, 
    \label{se}
\end{equation}
where 
\begin{equation}
    \mathcal{L}_e = \frac{1}{2}(\6_t a^i + \6^i a^0)^2  - \frac{c^2}{4} (f^{ij})^2~.
    \label{el}
\end{equation}
In the above $(a^0,a^i)$ are the vector potentials in Carroll limit which are related to their relativistic counter parts $A^\mu$ through the mappings given in (\ref{contrael}); i.e. $a^0 = A^0$ and $a^i = (1/c)A^i$.
Now varying the action (\ref{se}) with respect to $a^0, ~a^j$ we get the following equations respectively,
\begin{eqnarray}
    \6_i (\6_t a^i + \6^i a^0) = 0~, \label{ee1}\\
     \6_t (\6_t a^j + \6^j a^0) - c^2\6_i f^{ij} = 0~. \label{ee2}
    \end{eqnarray}

Now it is easy to show that these equations can be directly obtained from the relativistic equations of motion 
\begin{eqnarray}
\6_i F^{i0} =0~, \label{e1}\\
\6_0 F^{0j} + \6_i F^{ij} = 0~, \label{e2}
\end{eqnarray}
when one takes the Carroll limit. Use of our prescribed path on (\ref{e1}) for the electric limit yields
\begin{eqnarray}
\6_i F^{i0} =0  
\xrightarrow[\text{$c \to 0$}]{\text{electric~limit}} \6_i (\6_t a^i + \6^i a^0) = 0~,
\label{eo1}
\end{eqnarray}
which reproduces (\ref{ee1}). Likewise the Carrollian limit of (\ref{e2}) reproduces (\ref{ee2}).
This shows the consistency of the equations of motion in Carrollian Maxwell model.

\subsection{Magnetic limit}
\noindent In this limit the Maxwell action will be reduced to 
\begin{equation}
    S_m = \int d^4 x \mathcal{L}_m~, \label{sm}
\end{equation}
where
\begin{equation}
    \mathcal{L}_m = \frac{1}{2} \Big(\frac{1}{c}\6_t a^i + c\6^i a^0\Big)^2 - \frac{1}{4} \Big(f^{ij}\Big)^2~. \label{ml}
\end{equation}
In this case the relativistic fields and the Carrollian ones are mapped through (\ref{contramag}); i.e. $a^0=(1/c)A^0$ and $a^i=A^i$. 
Varying the action (\ref{sm}) with respect to $a^0, ~a^j$ we get the following equations respectively,
\begin{eqnarray}
    \6_i\6_t a^i = 0~, \label{em1} \\
    \6_t\Big(\frac{1}{c}\6_t a^j + c\6^i a^0\Big) - \6_i f^{ij} = 0~. \label{em2}
\end{eqnarray}
Now, like earlier, the above two equations can be directly obtained from (\ref{e1}) and (\ref{e2}) by imposing the magnetic limit -- (\ref{e1}) yields (\ref{em1}) while (\ref{em2}) is obtained from (\ref{e2}). This again verifies the consistency of our analysis. 
 %We now derive these equations directly from  relativistic equations of motion which can be written as
%\begin{eqnarray}
%\6_i F^{i0} =0 \label{m1}\\
%\6_0 F^{0j} + \6_i F^{ij} = 0 \label{m2}
%\end{eqnarray}
%From eqn \ref{m1} we get in the electric limit
%\begin{eqnarray}
%\6_i F^{i0} =0  
%\xrightarrow[\text{$c \to 0$}]{\text{magnetic~limit}} %\6_i \6_t a^i = 0
%\label{mo1}
%\end{eqnarray}
%which reproduces eqn \ref{em1}. Likewise the Carrollian limit of \ref{m2} reproduces \ref{em2}.

\section{Electric and magnetic fields in Carrollian Maxwell theory} \label{sec5}
The electric and magnetic fields at the Carroll limit and their mappings with the counter parts in relativistic regime can be obtained by using (\ref{contrael}) and (\ref{contramag} and the relativistic Maxwell's equations.
%\noindent \underline{\bf Relativistic Maxwell equations:}\\
The relativistic, source free Maxwell's equations in terms of electric and magnetic fields are given by
\begin{eqnarray}
    \vec  \nabla \cdot \vec E = 0~; \,\,\,\,\ 
    \vec\nabla  \cdot  \vec B = 0~; \nonumber \\
    \vec \nabla \times \vec E = -\frac{1}{c} \frac{\6 \vec B}{\6 t}~; \,\,\,\,\
    \vec \nabla \times \vec B = \frac{1}{c} \frac{\6 \vec E}{\6 t}~. \label{rmax}
\end{eqnarray}
Before going to the analysis we denote the electric and magnetic fields in Carroll limit as $e^i$ and $b^i$, respectively.
\textcolor{blue}{
\begin{table}
\caption{Field scalings in Galilean and Carrollian limit}\label{T05}
\begin{center}
\begin{tabular}{|c|c|c|} \hline 
${\rm Limits}$  & $ {\rm Galilean ~limit}$ & $ {\rm Carrollian ~limit}$  \\ \hline
${\rm Electric ~limit}$ & $E^i \to c e^i, ~B^i \to b^i$ & $E^i \to e^i, ~B^i \to cb^i$ \\ \hline
${\rm Magnetic ~limit }$ &  $ E^i \to \frac{e^i}{c}, ~B^i \to b^i$  & $E^i \to ce^i, ~B^i \to b^i $ \\
\hline
\end{tabular}
\label{T05}
\end{center}
\end{table}
}

\noindent \underline{\bf Electric limit:}\\
In the electric limit the mappings between the
relativistic and Carroll vector components are given by (\ref{contrael}). Therefore the four-vector potential components are related as $a^0 = A^0$ and $a^i = (1/c)A^i$. Now the second relation in (\ref{rmax}) yields $\vec{B} = \vec{\nabla}\times \vec{A}$; i.e. $B^i = \epsilon^{ijk}\partial^j A^k$. Then use of the mapping $A^i = ca^i$ shows that $B^i$ should go as $B^i = cb^i$. To identify the electric field it is reasonable to use the static situation of the Maxwell's equations. In this case one has $\vec{E} = - \vec{\nabla}A^0$ and as $A^0=a^0$, we must have $E^i=e^i$.     
Thus we have the following scaling relations:
\begin{equation}
    E^i \to e^i~; \,\,\,\ B^i \to c b^i~. \label{efmap}
\end{equation}
Using the above scaling relation and taking $c \to 0$ limit, the relativistic Maxwell's equations (\ref{rmax}) reduce to the following set of equations
\begin{eqnarray}
    \vec  \nabla \cdot \vec e = 0~; \,\,\,\,\
    \vec\nabla  \cdot  \vec b = 0~; \nonumber \\ 
     \vec \nabla \times \vec e = -\frac{1}{c} \frac{\6 \vec b}{\6 t}~; \,\,\,\,\
   \frac{\6 \vec e}{\6 t} = 0~. \label{eM}
\end{eqnarray}

\noindent \underline{\bf Magnetic limit:}\\
As in the magnetic limit the mappings for the vector potentials are $A^0=ca^0$ and $A^i=a^i$, use of the similar argument yields
the following scaling relations
\begin{equation}
    E^i \to ce^i~; \,\,\,\  B^i \to  b^i~. \label{mfmap}
\end{equation}
Using the above scaling relation and taking $c \to 0$ limit, the relativistic Maxwell's equations (\ref{rmax}) reduce to the following set of equations
\begin{eqnarray}
    \vec  \nabla \cdot \vec e = 0~; \,\,\,\,\
    \vec\nabla  \cdot  \vec b = 0~; \nonumber \\ 
     \frac{\6 \vec b}{\6 t} = 0~; \,\,\,\,\
     \vec \nabla \times \vec b = \frac{1}{c} \frac{\6 \vec e}{\6 t}~. \label{mM}
     \end{eqnarray}
Our results here (as given in (\ref{eM}) and (\ref{mM})) are consistent with the results discussed in \cite{Duval} which are derived in a different approach. One can also observe that a particular type of duality symmetry holds here. One can see from the set of equations given in (\ref{eM}) and (\ref{mM}) that under the transformation $e^i \to b^i$ and $b^i \to -e^i$ the electric limit equations (given in (\ref{eM})) transform to the magnetic limit equations (given in (\ref{mM})) and vice-versa. The results are summarised schematically in the box below.
\begin{center}
\fbox{\begin{minipage}{25em}
\begin{equation*}
    e^i \to \pm b^i, ~~~b^i \to \mp e^i \Longleftrightarrow {\rm {\bf Electric \longleftrightarrow Magnetic}}
\end{equation*}
\end{minipage}}
\end{center}

%\textcolor{red}{1. Comparison with earlier results?? 2. Comment on duality symmetry??}
     
\section{Transformations of fields under boost} \label{sec6}
So far we have considered the potentials and given the appropriate scaling in the electric and magnetic limits. Relevant Maxwell's equations in the Carroll limit were obtained. To derive the scaling relations for the fields we considered the definition of the fields in terms of the potentials. Next, exploiting the scaling relations for potentials we can obtain the corresponding relations for fields. This is shown below.

The field tensor transforms as
\begin{eqnarray}
F'^{\mu \nu}(x') = \frac{\6 x'^{\mu}}{\6 x^{\lambda}}\frac{\6 x'^{\nu}}{\6 x^{\rho}} F^{\lambda \rho}(x)~. 
\label{def1}
\end{eqnarray}
The Sengupta's form of boost transformations are given by (\ref{t1}) and (\ref{t2}).
%\begin{equation}
%    x'^0 = \tilde\gamma x^0 - \tilde\gamma \tilde\beta_i x^i
%    \label{t6}
%\end{equation}
%\begin{equation}
%    x'^i= \tilde \gamma x^i - \tilde\gamma \tilde\beta^i x^0 
%    \label{t7}
%\end{equation}
Using these in (\ref{def1}) we get following relations
\begin{eqnarray}
&&F'^{0i} = E'^i = \tilde \gamma E^i + \tilde \gamma \Big(\tilde \gamma -1 \Big) \frac{\tilde \beta^i \tilde \beta_j}{\tilde \beta^2} E^j - \tilde \gamma^2 \tilde \beta^i \tilde \beta_j E^j  + \tilde\gamma \tilde\beta_j F^{ij} + \tilde \gamma \Big(\tilde \gamma -1  \Big) \frac{\tilde \beta^i \tilde \beta_j \tilde \beta_m}{\tilde \beta^2} F^{jm}~; \label{e}
\\
&&F'^{ij} 
=  -\tilde\gamma \tilde\beta^i E^j+ \tilde\gamma \tilde\beta^j E^i + F^{ij} + \Big( \tilde \gamma -1 \Big) \frac{\tilde \beta^j \tilde \beta_k}{\tilde \beta^2} F^{ik} + \Big( \tilde \gamma - 1\Big) \frac{\tilde \beta^i \tilde \beta_l}{\tilde \beta^2} F^{lj} + \Big(\tilde \gamma -1  \Big)^2 \frac{\tilde \beta^i \tilde \beta^j \tilde \beta_l \tilde \beta_k}{\tilde \beta^4} F^{lk}~. \label{b}
\end{eqnarray}
Now we are going to discuss two sectors of the Carroll limit by taking $c \to 0$ limit, more precisely imposing (\ref{B5}). 
%Now in this limit we know $\tilde \gamma \to 1$ and $\tilde \beta^i << 1$.

\noindent \underline{\bf Electric limit:}\\
\noindent Here using the electric limit scaling relation (\ref{efmap}) in Eqs. (\ref{e}) and (\ref{b}) we get the following field transformations equations under Carrollian boost:
\begin{eqnarray}
    e'^i = e^i~; \\
   b'^i = b^i - \frac{1}{\omega^2} \Big(\vec \omega \times \vec e \Big)^i~.
\end{eqnarray}

\noindent \underline{\bf Magnetic limit:}\\
\noindent Here using the magnetic limit scaling relation (\ref{mfmap}) in Eqs. (\ref{e}) and (\ref{b}) we get the following field transformations equations under Carrollian boost:
\begin{eqnarray}
    e'^i = e^i + \frac{1}{\omega^2} \Big( \vec \omega \times \vec b  \Big)^i~; \\
    b'^i = b^i~.
\end{eqnarray}
A comparison with the Galilean boost transformations is provided in table \ref{T04}
\begin{table}
\caption{Transformation of fields under Galilean and Carrollian boost}\label{T04}
\begin{center}
\begin{tabular}{|c|c|c|} \hline 
${\rm Limits}$  & $ {\rm Galilean~case}$ & $ {\rm Carroll~case}$  \\ \hline
${\rm Electric ~limit}$ & $e'^i = e^i, ~~b'^k =  b^k -\Big(\vec v \times \vec e \Big)^k$ & $e'^i = e^i, ~~b'^i = b^i - \frac{1}{\omega^2} \Big(\vec \omega \times \vec e \Big)^i$\\ \hline
${\rm Magnetic ~limit }$ &  $e'^i = e^i + \Big(\vec v \times \vec b \Big)^i, ~~\vec b' = \vec b $  & $e'^i = e^i + \frac{1}{\omega^2} \Big( \vec \omega \times \vec b  \Big)^i, ~~b'^i = b^i$ \\
\hline
\end{tabular}
\label{T04}
\end{center}
\end{table}\\
\noindent  It is interesting to note that the scaling relations for the fields may be derived directly using boost transformations. For the Galilean electrodynamics a similar exercise has been done in \cite{Rousseaux2}. The appropriate boost transformations are reproduced for the scaling relations given in table \ref{T05}. This provides an additional justification for taking those relations. Now to proceed further one has to compare the module of the electric field with that of the magnetic field. So for the {\it electric limit}, where electric field is dominating than the magnetic field, we get
\begin{eqnarray}
    |E| >> |B|; \label{emod1}~\\
     e'^i = e^i~; \label{emod2}\\
   b'^i = b^i - \frac{1}{\omega^2} \Big(\vec \omega \times \vec e \Big)^i\label{emod3}~.
\end{eqnarray}
One can clearly see from the electric limit condition \ref{emod1} that scaling relation provided in the first row of table \ref{T05} is justified in the $c \to 0$ limit. Similarly for the {\it magnetic limit} we get,
\begin{eqnarray}
     |E| << |B|; \label{bmod1}~\\
      e'^i = e^i + \frac{1}{\omega^2} \Big( \vec \omega \times \vec b  \Big)^i~; \\
    b'^i = b^i~.
\end{eqnarray}
Again the scaling relation given in the second row of table \ref{T05} is justified from the magnetic limit condition given in eqn \ref{bmod1}. From the above relations one can derive the transformations of the potentials in both electric and magnetic limits.

\section{Gauge symmetry} \label{sec7}
\noindent We know in the relativistic case the Maxwell action, given by (\ref{mlag}) along with (\ref{B6}),
%\begin{equation}
%  \mathcal{L} = -\frac{1}{4} F_{\mu \nu} F^{\mu \nu} = -\frac{1}{4}\Big(\eta_{\mu \lambda} \eta_{\nu \rho} F^{\lambda \rho} F^{\mu \nu}   \Big) 
%\end{equation}
is invariant under the following gauge transformation,
\begin{equation}
   \delta A^{\mu} = \6^{\mu} \alpha~, \label{cpot}
\end{equation}
where $\alpha$ is an arbitrary function of spacetime.
Now we consider two sectors of the Carroll limit.

\noindent \underline{\bf Electric limit:}\\
In electric limit we deduce following relations:
\begin{eqnarray}
\delta A^0 = \6^0 \alpha \implies \delta a^0 = -\frac{1}{c} \6_t \alpha~;  \label{eg3}
\end{eqnarray}
and
\begin{eqnarray}
\delta A^i = \6^i \alpha \implies c \delta a^i = \6^i \alpha \implies \delta a^i = \frac{1}{c} \6^i \alpha~. \label{eg4}
\end{eqnarray}
Now one can verify that the variation of the action (\ref{el}) under (\ref{eg3}) and (\ref{eg4}) vanishes:
\begin{eqnarray}
    \delta \mathcal{L}_e = \Big(\6_t a^i + \6^i a^0\Big)\Big( \6_t \delta a^i + \6^i \delta a^0  \Big)-\frac{c^2}{2} (f^{ij})(\delta f^{ij}) \nonumber \\
    = \Big(\6_t a^i + \6^i a^0\Big) \Big(\frac{1}{c} \6_t \6^i \alpha - \frac{1}{c} \6^i \6_t \alpha    \Big) -\frac{c^2}{2} (f^{ij}) \Big(\frac{1}{c} \6^i \6^j \alpha - \frac{1}{c}\6^j \6^i \alpha     \Big) = 0~.
\end{eqnarray}
Hence (\ref{eg3}) and (\ref{eg4}) define the Carrollian gauge transformations in the electric limit.

\noindent \underline{\bf Magnetic limit:}\\
In magnetic limit we deduce following relations:
\begin{eqnarray}
&&\delta A^0 = \6^0 \alpha \implies c\delta a^0 = -\frac{1}{c} \6_t \alpha \implies \delta a^0 = -\frac{1}{c^2} \6_t \alpha~; \label{mg1}
\\
&&\delta A^i = \6^i \alpha \implies  \delta a^i = \6^i \alpha~.  \label{mg2}
\end{eqnarray}
Like earlier, here also the action (\ref{ml}) is invariant under these transformations: 
\begin{eqnarray}
    \delta \mathcal{L}_m =  \Big(\frac{1}{c}\6_t a^i + c\6^i a^0\Big)\Big(\frac{1}{c}\6_t \delta a^i + c\6^i \delta a^0\Big) - \frac{1}{2} \Big(f^{ij}\Big) \Big(\delta f^{ij}\Big)\nonumber \\
    = \Big(\frac{1}{c}\6_t a^i + c\6^i a^0\Big) \Big(\frac{1}{c}\6_t \6^i \alpha  - \frac{1}{c}\6^i \6_t a^0\Big) - \frac{1}{2} \Big(f^{ij}\Big) \Big( \6^i \6^j \alpha - \6^j \6^i \alpha     \Big) = 0~. 
\end{eqnarray}
Therefore (\ref{mg1}) and (\ref{mg2}) construct the required gauge in the magnetic limit.

\subsection{Noether current conservation}
In relativistic classical field theory the Noether current corresponding to the symmetry transformation $A_\mu\to A_\mu+\delta A_\mu$ is given by
\begin{eqnarray}
 J^{\mu}= \frac{\6 \mathcal{L}}{\6 (\6_\mu A_\nu)} \delta A_\nu~. \label{current1}
\end{eqnarray}
which is conserved on-shell i.e $\6_\mu J^\mu =  0$. Note that here $\delta A_\mu$ corresponds to a  completely arbitrary symmetry transformation. Specifically for the gauge transformation, given by (\ref{cpot}), the current takes the form 
\begin{equation}
    J^\mu = -F^{\mu \nu} \6_\nu \alpha~. 
\end{equation}
Using the equation of motion $\6_\mu F^{\mu \nu}= 0 $ one can easily show the conservation of the above current.

\noindent \underline{\bf Electric limit:}\\
Using the map given in (\ref{contrael}) we can have the electric Carrollian currents of the following forms
\begin{eqnarray}
    j^0 = \Big(\6_t a^i + \6^i a^0   \Big) \6_i \alpha~; \\
    j^i = -\frac{1}{c^2} \Big(\6_t a^i + \6^i a^0   \Big) \6_t \alpha - f^{ij} \6_j \alpha~. 
\end{eqnarray}
We can show the on-shell conservation of the electric Carrollian currents as follows: 
\begin{eqnarray}
\6_\mu J^\mu \xrightarrow[]{\text{Electric limit}} \frac{1}{c} \6_t j^0 + c \6_i j^i \xrightarrow[]{\text{$c \to 0$}}  0~.
\end{eqnarray}
In the above at the last stage equations of motion (\ref{ee1}) and (\ref{ee2}) have been used.

\noindent \underline{\bf Magnetic limit:}\\
Using the map given in (\ref{contramag}) we can have the magnetic Carrollian currents as
\begin{eqnarray}
    j^0 = \Big(\frac{1}{c^2} \6_t a^i + \6^i a^0  \Big) \6_i \alpha~; \\
    j^i = -\frac{1}{c} \Big( c \6^i a^0 + \frac{1}{c} \6_t a^i  \Big) \6_t \alpha - f^{ij} \6_j \alpha~.
\end{eqnarray}
In a similar manner using (\ref{em1}) and (\ref{em2}) we can show the on-shell conservation of the magnetic Carrollian current:
\begin{eqnarray}
\6_\mu J^\mu \xrightarrow[]{\text{Magnetic limit}}  \6_t j^0 +  \6_i j^i \xrightarrow[]{\text{$c \to 0$}}  0~.
\end{eqnarray}

\section{Shift symmetry} \label{sec8}
\noindent We know that Goldstone's theorem is a crucial input of the study of low-energy effective lagrangians implying that whenever a global symmetry is spontaneously broken, a massless mode will appear. In relativistic theories this leads to a massless Goldstone particle described by a shift symmetry of
the field
\begin{equation}
    \phi(x) \to \phi(x) + \mathcal{C}~, \label{fi}
\end{equation}
where $\mathcal{C}$ is constant and is characterised by the scalar field action
\begin{equation}
    S = \frac{1}{2} \int d^d x \6_\mu \phi \6^\mu \phi~.
\end{equation}
The above action is invariant under (\ref{fi}). Since (\ref{fi}) is a global transformation, the conserved current can be found by exploiting Noether's first theorem. This is given by 
\begin{eqnarray}
J^\mu = \frac{\6 \mathcal{L}}{\6(\6_\mu \phi)} \delta \phi = \mathcal{C}\6^\mu \phi~,
\end{eqnarray}
and the corresponding conservation is demonstrated as 
\begin{equation}
    \6_\mu J^\mu = \mathcal{C} \6_\mu \6^\mu \phi = 0~.
\end{equation}
The last equality is obtained by using the equation of motion for the scalar field.
Now comeback to the Maxwell theory. In this case the action (\ref{mlag}) is invariant under a constant shift in the four potential: 
\begin{equation}
    A'^{\mu} = A^{\mu} + C^{\mu}~.
\end{equation}
Then the conserved current (\ref{current1}) reduces to
\begin{equation}
J^{\mu}= \frac{\6 \mathcal{L}}{\6 (\6_\mu A_\nu)} C_\nu~.
\label{B8}
\end{equation}

\noindent \underline {\bf Electric limit} \\
We can define following things
\begin{eqnarray}
 \delta A^0 = C_0 \implies \delta a^0 = C^0~;  \\
\delta A^i = C_i \implies c \delta a^i = C^i \implies \delta a^i = \frac{1}{c} C^i~.
\end{eqnarray}
From (\ref{el}) the components of Noether current are found to be
\begin{eqnarray}
j^0 = \Big(\6_t a^i + \6^i a^0  \Big) C^i~; \,\,\, j^i =  -\frac{1}{c} \Big(\6_t a^i + \6^i a^0 \Big) C_0 + f^{ij} C^j~.
\end{eqnarray}
The current conservation can be explicitly demonstrated as 
\begin{eqnarray}
\6_\mu J^\mu \xrightarrow[]{\text{Electric limit}} \frac{1}{c} \6_t j^0 + c \6_i j^i \xrightarrow[]{\text{$c \to 0$}}  0~.
\end{eqnarray}

\noindent \underline {\bf Magnetic limit} \\
Now We have the following things
\begin{eqnarray}
 \delta A^0 = C^0 \implies c \delta a^0 = C^0 \implies \delta a^0 = \frac{1}{c} C^0~;  \\
\delta A^i = C_i \implies  \delta a^i = C^i~. 
\end{eqnarray}
From (\ref{ml}) the components of Noether current are found to be
\begin{eqnarray}
    j^0 = \Big( \frac{1}{c} \6_t a^i + c \6^i a^0  \Big)~; \,\,\,\,\ j^i = \Big(c \6^i a^0 + \frac{1}{c} \6_t a^i    \Big) C^i - f^{ij} C^j
\end{eqnarray}
and corresponding current conservation can be demonstrated as 
\begin{eqnarray}
\6_\mu J^\mu \xrightarrow[]{\text{Magnetic limit}}  \6_t j^0 +  \6_i j^i \xrightarrow[]{\text{$c \to 0$}}  0~.
\end{eqnarray}

\section{Conclusions} \label{sec9}
\noindent Let us now summarise the new significant findings of the paper, comparing with existing results found in the literature.\\
\indent Non-Lorentzian physics, especially the Carroll type, deserve further study. For example, nowhere a systematic construction of an action principle was presented till now. Here in this paper we put forward a formulation to study the Carrollian physics based on Sengupta's transformation \cite{Sengupta} which is complementary to the standard Lorentz transformations. We show that using these transformations one can systematically build up an action formulation for Carrollian Maxwell theory. A central point of this paper is the formulation of a consistent dictionary that translates Lorentzian four vectors in the relativistic theory to their Carrollian counterpart. \\
    \indent  Based on this action principle, the equations of motion are derived both for potential and fields. Their internal consistency has been checked thoroughly. We observe a duality symmetry between Carrollian electric and magnetic sector. Explicit demonstration of boost invariance has been presented.\\
    \indent  Also, we know gauge symmetries and shift symmetries play a pivotal role in the
 understanding of different gauge theories. Since relativistic Maxwell theory possess both gauge and shift symmetry,  it is interesting to observe its consequence in the Carrollian invariant theory which has been derived by using the dictionary presented here. We have extensively shown here the corresponding current conservations for both the symmetries. 

Carrollian electrodynamics has been discussed earlier as well. However our approach is different from these attempts not only from the construction point of view, but also in terms of the robustness of the physical arguments. In \cite{Duval}, the same has been constructed through a geometrical point of view. However it did not introduce the action in the Carrollian limit. \cite{Mehra} also briefly mentioned about the same in the light of flat space holography, however without any systematic analysis. Their discussion does not provide the physical construction of the underlying scaling nature of the potential vector components and therefore carries a possible ambiguity in the analysis. Although \cite{Boar} captures the action formulation of the Carroll electrodynamics, but their approach to construct the same is different. We have started from Sengupta transformation to motivate the scaling behaviour of the Carroll vectors while \cite{Boar} considered the Lorentz transformations as the guiding one. They have expanded the Lorentz transformation in powers of $c$ to obtain the required scaling laws. A similar expansion method on the potential vector has been adopted in \cite{Minhaz} to derive the Carroll limit of Yang-Mills theory. In contrary the Sengupta transformations played the pivotal role in present analysis and it helped to obtain a systematic construction staring from the action formulation of Carrollian electrodynamics. In this sense it provides a much robust and physically motivated formulation. Galilean transformations are non-relativistic limit of Lorentz transformations. Analogously we derived here the Carroll transformations from the Sengupta ones.  

\noindent \underline{\bf Future prospects:}\\

\noindent Given the importance of the Carrollian limit, which plays a ubiquitous role in different arena of physics, following are the issues that we want to address in near future.\\
%\begin{itemize}
\indent In the present paper we mostly focus on Maxwellian electrodynamics. It is interesting to use the present formalism to other gauge theories for example Proca model, Born-Infeld theory, Chern-Simons theory etc. Some works have been done in this direction in recent time but we believe they are far from complete. At least none of these works provided any systematic action formalism. The Hamiltonian formalism in the context of Carrollian relativistic theory is also another aspect that we want to look at. \\
\indent Carroll limit plays important role in different,  gravitational aspects particularly in the context of gravitational waves and memory effects \cite{Duval3, Duval4}. We would like to do some further study of memory effect from Carroll symmetry point of view. We believe our study will shed some important light on the unknown corners.\\
\indent We know that ever since its inception, there have been numerous attempts to extend the original holographic AdS/CFT correspondence to include asymptotically flat spacetime or de-Sitter spacetime. Fluid/gravity correspondence have been considered as a promising way out to reach final construction of flat space holography. The non-relativistic hydrodynamics play an important role due to the incomplete understanding on the role played by the null infinity. It is well established that the stretched horizon in the membrane paradigm is described by Galilean hydrodynamics \cite{Damour}. However it is recently shown in \cite{Petkou} that the boundary fluids which are holographically dual to Ricci-flat spacetimes, are described by Carrollian hydrodynamics. Hence both {\it Galilean} and {\it Carrollian} limit play important role \cite{Petkou, Petkou2, Donnay}. We want to look into both Galiean and Carrollian aspects of hydrodynamics in near future. 
%\end{itemize}
\section*{Acknowledgements}
Two of the authors (RB and SB) acknowledge the support from a DAE Raja Ramanna Fellowship (grant no: 1003/(6)/2021/RRF/R$\&$D -- II/4031, dated: 20/03/2021). The other author (BRM) is supported by Science and Engineering Research Board (SERB), Department of Science $\&$ Technology (DST), Government of India, under the scheme Core Research Grant (File no. CRG/2020/000616).

\end{document}